# MICROLENSING INDUCED SPECTRAL VARIABILITY IN Q2237+0305

GERAINT F. LEWIS

*Institute of Astronomy, Cambridge. UK*

MIKE J. IRWIN

*Royal Greenwich Observatory, Cambridge. UK*

AND

PAUL C. HEWETT

*Institute of Astronomy, Cambridge. UK*

## 1. Introduction

Several macrolensed systems exhibit photometric variability consistent with microlensing due to objects of stellar mass located in the lens. The degree of microlensing amplification is dependent upon the size of the source, with smaller sources being more amplified. In general, amplification of sources larger than an Einstein radius projected onto the source plane is negligible. For the quasar Q2237+0305, a quadruple–image lens (Huchra *et al.*, 1985), this radius is 0.05 pc, larger than the predicted size of a continuum–emitting accretion disk, but substantially smaller than the broad line region (Figure 1). This scale difference implies that the continuum will be amplified while the broad line emission remains essentially unchanged during a microlensing event (Sanitt, 1971; Kayser *et al.*, 1986).

The broad line emitting region, as a whole, is too large to be microlensed, but substructure on small scales may be significantly amplified. Although the total flux in the line is relatively unchanged, microlensing of substructure can result in changes in the shape of the emission line profiles, and produce measurable shifts in the central wavelength of the line (Nemiroff, 1988; Schneider and Wambsganss, 1990).

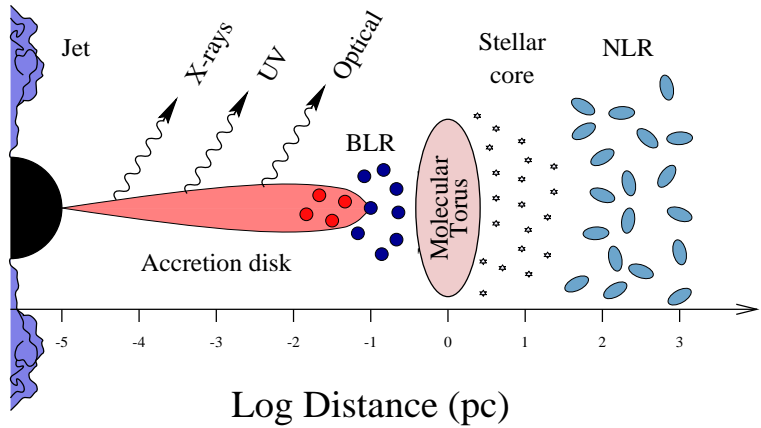

Figure 1. A schematic representation of the scales of structure at the centre of active galactic nuclei (Rees, 1984).

## 2. Observations and Spectral Modelling

Data were acquired at two epochs, 1991 August 14/15 and 1994 August 17/18, using the William Herschel Telescope at the Roque de los Muchachos Observatory, La Palma. The telescope was equipped with the ISIS double–beam spectrograph, providing a wavelength coverage of $\lambda\lambda 4000 - 8000$, including the emission lines of CIV $\lambda 1549$, CIII] $\lambda 1909$ and MgII $\lambda 2798$ in the quasar. The spectrograph slit was oriented to acquire spectra of pairs of images simultaneously. Direct $R$–band images were acquired at the Cassegrain auxiliary port, allowing the determination of the magnitudes of the four quasar components (Figure 2).

The spectroscopic CCD frames were debiased, flat fielded and sky subtracted using standard techniques. The resulting frames consisted of pairs of quasar spectra, separated by $\sim 1''$, superposed upon the extended background of the lensing galaxy light. To extract the individual quasar spectra, a model was constructed to describe the spatial distribution of light at each pixel in the dispersion direction. The model consisted of two point spread functions (the quasars) and an extended profile to represent the galaxy light. Figure 3 presents an example of the model fit to a single spatial cut across the CCD frame.

## 3. Results

Figures 4 and 5 present the spectra of pairs of images, (A+C, A+D, B+C, B+D), from the 1991 observations. In each case, the spectrum of the fainter image (C or D) is scaled such that its continuum matches the brighter image (A or B). Excess emission line flux of the scaled spectrum over that of the other image is indicated by the solid black regions.

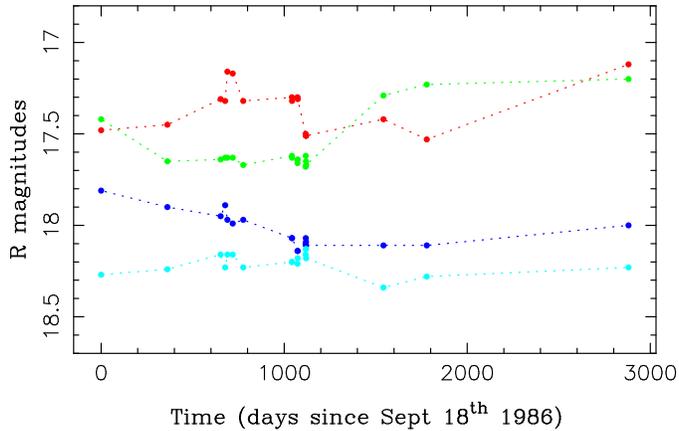

*Figure 2.* The $R$–band light curve for the images of Q2237+0305. The magnitudes are from the work of Houde and Racine (1994), the HST data of Rix *et al.* (1992) and our own observations (last two epochs).

The photometry time series (Figure 2) shows that during the 1991 observations images A and C were faint in comparison to their previous behaviour. The top panels in Figure 4 reveal that, when scaled so that the continuum levels match, the A and C components possess very similar line strengths in all three emission lines visible.

At the same epoch image D was faint. When scaled and compared to the spectrum of image A (lower panels of Figure 4) image D displays additional flux in all three lines. This can be interpreted as a microlensing deamplification of the continuum in image D.

In 1991 image B was bright, possibly undergoing microlensing, suggesting an enhancement of the continuum should be evident. This is confirmed

| Epoch | A | B | C | D |
|-------|------|------|------|------|
| 1991  | 1.10 | 0.64 | 1.00 | 1.73 |
| 1994  | 0.70 | 0.70 | 1.00 | –    |

TABLE 1. The equivalent widths of the CIV line. At each epoch the values have been normalized with respect to image C.

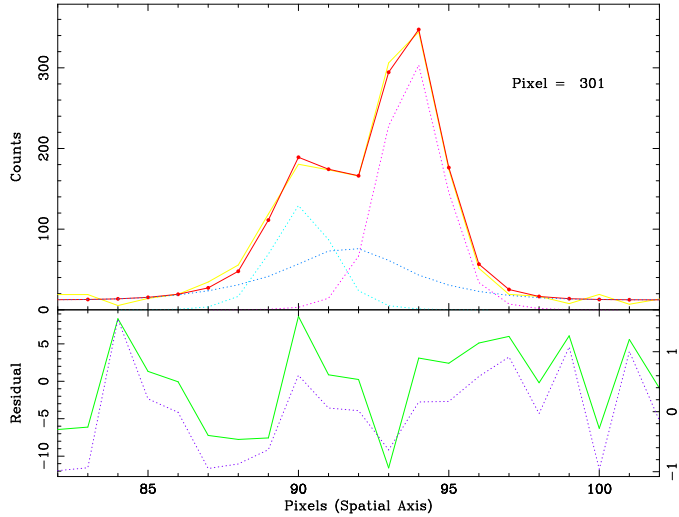

*Figure 3.* Spatial cross–section from a typical CCD exposure showing the model fit to the data. The model includes three components, two quasars and the galaxy (dashed lines). In the wings, at pixel positions less than 86 and greater than 98, the model consists of a contribution from the galaxy profile only. The $\chi^2$–statistic indicates the model adequately describes the data. The lower panel shows the residuals model − data, in counts (solid line) on the left hand axis, and in counts divided by the noise at each pixel position (dashed line) on the right hand side.

from the comparison between the scaled spectra of images B and C (top panels of Figure 5).

Table 1 presents the relative equivalent width of the CIV emission line, at both epochs, for all four components. The values are normalized with respect to component C. In 1991 the equivalent width measures confirm the impression gained from Figures 4 and 5, with very similar values for images A and C. Values for 1994 show that image B remained essentially unchanged while the equivalent width in image A dropped substantially.

Cross–correlation of the emission lines in the different components show no detectable velocity shifts in the line centroids. Within the limits of the S/N there are also no detectable emission line profile differences. These results indicate that the scale of any structure in the broad line emitting region must be greater than an Einstein radius projected into the source plane ($> 0.05\,pc$). Details of these results will appear in an article currently in preparation.

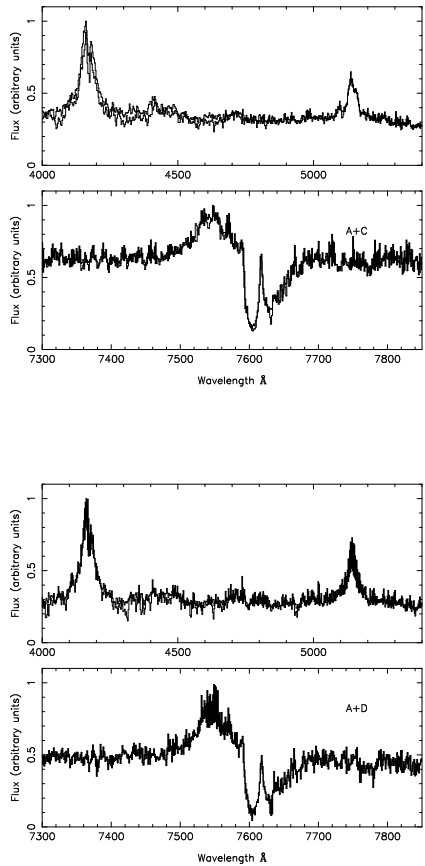

*Figure 4.* Blue and red ISIS spectra for images A + C (top panel) and A + D (bottom panel) from the 1991 observations. The spectra of C and D are scaled such that their continua match that of image A. The black regions indicate the excess of the scaled spectra over that of image A.

## 4. Conclusions

This paper shows evidence for equivalent width differences between the four images of Q2237+0305, providing the first spectroscopic detection of microlensing induced spectral variations in a quasar. The temporal variation of the equivalent widths are consistent with the microlensing hypothesis and indicate that the continuum in quasars originates in a region small, compared to the scale of an Einstein radius, while the broad line emission emanates from a much larger volume, in accord with the predictions of the standard model for the central regions of active galaxies (Figure 1).

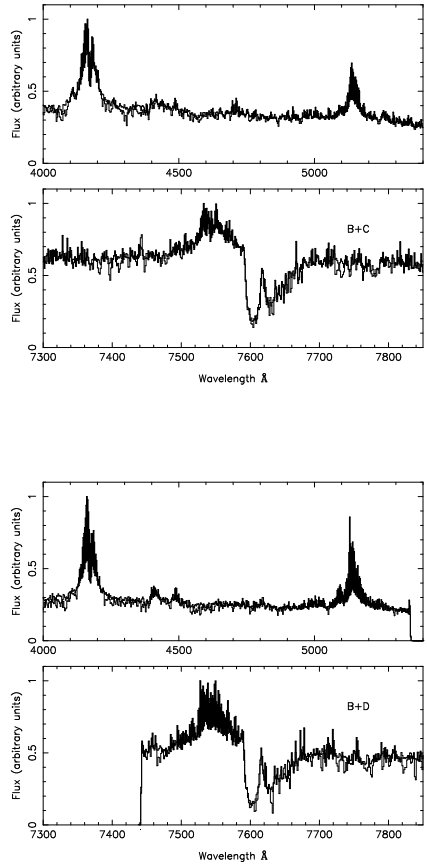

*Figure 5.* As for Figure 4, except for images B + C (top panel) and B + D (bottom panel).